# Who Owns the Knowledge? Copyright, GenAI, and the Future of Academic Publishing

Dmitry Kochetkov[1,2]
[1]Centre for Science and Technology Studies (CWTS), Leiden University, Leiden, the Netherlands
[2]Institute for the Study of Science, Russian Academy of Sciences, Moscow, Russia

E-mail: d.kochetkov@cwts.leidenuniv.nl
ORCID: 0000-0001-7890-7532

**Abstract.** The integration of generative artificial intelligence (GenAI) and large language models (LLMs) into scientific research and higher education presents a paradigm shift, offering revolutionizing opportunities while simultaneously raising profound ethical, legal, and regulatory questions. This study examines the complex intersection of AI and science, with a specific focus on the challenges posed to copyright law and the principles of open science. The author argues that current regulatory frameworks in key jurisdictions like the United States, China, the European Union, and the United Kingdom, while aiming to foster innovation, contain significant gaps, particularly concerning the use of copyrighted works and open science outputs for AI training. Widely adopted licensing mechanisms, such as Creative Commons, fail to adequately address the nuances of AI training, and the pervasive lack of attribution within AI systems fundamentally challenges established notions of originality. While current doctrine treats AI training as potentially fair use, this paper argues such mechanisms are inadequate and that copyright holders should retain explicit opt-out rights regardless of fair use doctrine. Instead, the author advocates for upholding authors' rights to refuse the use of their works for AI training and proposes that universities assume a leading role in shaping responsible AI governance. The conclusion is that a harmonized international legislative effort is urgently needed to ensure transparency, protect intellectual property, and prevent the emergence of an oligopolistic market structure that could prioritize commercial profit over scientific integrity and equitable knowledge production. This is a substantially expanded and revised version of a work originally presented at the 20th International Conference on Scientometrics & Informetrics (Kochetkov, 2025).

## Introduction

While artificial intelligence (AI) research dates back to the 1956 Dartmouth Conference (Strickland, 2021), recent advances in deep learning and natural language processing (NLP) (Vaswani et al., 2017) have enabled large language models (LLMs) capable of processing and generating human-like content at unprecedented scale. These capabilities have particular significance for scientific publishing, where LLMs are increasingly deployed for tasks ranging from peer review assistance (Zhang & Abernethy, 2025; Zhuang et al., 2025) to automated literature reviews (De La Torre-López et al., 2023).

The development of AI technology presents both challenges and opportunities across various fields (Rama Padmaja & Lakshminarayana, 2024; Wolff et al., 2018). While AI offers immense potential, its advancement raises ethical concerns, including biases, privacy issues, and broader social implications (Rama Padmaja & Lakshminarayana, 2024). AI's influence spans all five dimensions of sustainability, with both positive and negative consequences (Khakurel et al., 2018). For instance, an analysis of a Google Scholar sample of questionable scientific papers suspected to be generated by GPT revealed that many address applied, often controversial issues prone to misinformation, such as environment, health, and computing (Haider et al., 2024).
AI is transforming research jobs, and in turn, science provides LLMs with a vast amount of data for training. However, LLMs may pose a threat to copyright, as they can generate content that potentially violates intellectual property rights (German, 2024). Currently, neither copyright nor "open" licenses can protect scholarly content from without author consent reuse in AI training (Decker, 2025). This fact raises fundamental questions that existing legal frameworks struggle to address.
The intersection of AI training and copyright law has generated substantial legal scholarship examining how different jurisdictions approach these challenges. Sag & Yu (2024) identify an emerging international equilibrium where countries recognize text and data mining as potentially valuable while maintaining some copyright protections, driven by the centrality of the idea-expression distinction, global AI competition, and reform convergence. Chopra et al., (2025) offer comparative analysis of how courts and policymakers in multiple jurisdictions address generative AI's impact on copyright and personality rights, highlighting divergent regulatory approaches between U.S. litigation-based doctrines, EU statutory frameworks, and emerging jurisprudence in other legal systems. The European Parliament's 2025 study on generative AI and copyright examines structural risks to European creative economy, calling for harmonized opt-out mechanisms and equitable licensing models (Lucchi, 2025). Senftleben (2025) proposes policy pathways to balance innovation with creator protection.
This study examines how generative AI and LLMs challenge existing copyright frameworks in scholarly publishing, an issue that remains underdeveloped in current policy discussions. Within the broader discussion about AI's impact on science, I focus specifically on three interconnected issues: (1) whether current licensing mechanisms adequately protect scholarly works from unauthorized AI training, (2) how regulatory frameworks across major jurisdictions address these copyright concerns, and (3) what actions stakeholders can take to establish fairer governance of AI training on academic content. The focus on these three issues responds to an urgent gap in current policy discussions, where the use of publicly available research outputs for training large language models remains largely unregulated.

## AI-Related Regulations

In this section, I provide an analysis of the regulations related to artificial intelligence (AI) in the United States, China, the United Kingdom, the European Union, and major international initiatives influencing national AI legislation.

### *United Kingdom*

Interestingly, there is currently no comprehensive regulation governing AI in the UK and the United States. The Sunak government issued a framework document in 2023 titled *A Pro-Innovation Approach to AI Regulation* (2023), which establishes basic principles for AI. The document promotes flexible regulation and aims to foster innovation through the development and use of AI technologies. The British government has also expressed its ambition to make the UK the best place to invest in AI.
The Artificial Intelligence (Regulation) Bill, reintroduced as a Private Members' Bill by Lord Holmes of Richmond on March 4, 2025, represents a significant departure from the government's earlier voluntary approach by proposing the creation of a centralized AI Authority to coordinate regulatory oversight across sectors, establish regulatory sandboxes for testing AI innovations with real consumers, and mandate that all organizations developing, deploying, or using AI designate AI officers and undergo independent audits by accredited third parties. The bill specifically addresses the copyright protection regulatory gap by requiring organizations involved in AI training to report all third-party data and intellectual property used to the AI Authority with assurances of informed consent and compliance with copyright laws, alongside mandatory health warnings, labeling, and consent options for AI products and services, thus providing the transparency and accountability mechanisms around training data provenance that have been largely absent from other AI regulatory frameworks.
The Data (Use and Access) Act 2025, which received Royal Assent on June 19, 2025, represents a foundational intervention: rather than establishing immediate copyright protection, it creates a statutory framework for future copyright and AI training governance. The Act does not directly prohibit unlicensed AI training or establish opt-out rights; instead, it imposes government obligations to evaluate whether such protections are needed and to report recommendations to Parliament. The UK government has established three competing policy objectives that any copyright-AI framework must balance – (1) control (rightsholders' ability to license and monetize), (2) access (developers' lawful access to training data), and (3) transparency (clear framework with disclosure of training sources). At the same time, the Act involves Secretary of State obligations (due March 18, 2026): (1) economic impact assessment of all policy options; (2) comprehensive report on use of copyright works in AI development.
The December 2025 progress statement indicates the four policy options under consideration:

- Option 0: do nothing
- Option 1: strengthen copyright – requiring licensing in all cases
- Option 2: a broad data mining exception[1]
- Option 3: a data mining exception with rights reservation underpinned by supporting measures on transparency (Government's preferred option).

Option 3 assumes default permission for AI training unless copyright holder expressly reserves rights (opt-out mechanism). It looks like a balanced approach but requires technical opt-out standards. Notably, the consultation received over 11,500 responses, with 88% supporting mandatory licensing (Option 1) versus only 3% supporting the government's preferred opt-out framework (Option 3). It looks like a public mandate for author-protective policy that may fundamentally reshape the UK's regulatory trajectory.

*United States*

In the United States, a framework document was published in October 2023, titled *Executive Order 14110 on the Safe, Secure, and Trustworthy Development and Use of Artificial Intelligence* (2023). Notably, this document included actions related to copyright law, stating: "…consult with the Director of the United States Copyright Office and issue recommendations to the President on potential executive actions relating to copyright and AI. The recommendations shall address any copyright and related issues discussed in the United States Copyright Office's study, including the scope of protection for works produced using AI and the treatment of copyrighted works in AI training."
A significant step forward was taken with the development of the *Generative AI Copyright Disclosure Act of 2024* (Schiff, 2024). This act aimed to ensure transparency in the use of copyrighted works for AI training and is currently under consideration in the House of Representatives. If passed, the act would require companies to notify the U.S. Copyright Office about any copyrighted works used in their AI systems. These notifications must be submitted 30 days before or after the public release of the AI system, ensuring transparency and accountability. The act is intended to help copyright holders make informed decisions about licensing and compensation. However, the wording of the document remains vague, raising questions for both AI developers and copyright owners. Additionally, I have concerns about the inability of copyright holders to prohibit the use of their works for AI training, which creates a bias in favor of bigtech AI development.
The future of this Act is unclear because Trump rescinded Executive Order 14110 during the first days of his presidency. His *Executive Order 14179*, signed on January 23, 2025, articulated the need to develop a new approach to AI development (Executive Order 14179 Removing Barriers to American Leadership in Artificial Intelligence, 2025). This approach was formulated in *AI Action Plan* (2025) that prioritize "American values" like free speech in AI development, support open models, streamline permitting for infrastructure, support American workers, and build global alliances on AI standards and security. However, I didn't find any mentions of copyright in the document.
AI Action Plan was accompanied by a number of executive orders, namely:

- Promoting the export of the American AI technology stack
- Accelerating federal permitting of data center infrastructure
- Preventing "Woke AI" in the Federal Government

[1] Statutory exception permitting all AI training without copyright holder consent; no opt-out mechanism; strongest AI developer protection; weakest author protection.

The documents signed on July 23, 2025, clearly indicate that AI, like many other spheres of American social life under Trump's administration, proved to be highly politicized.
One of the most significant recent developments in state-level AI regulation is the California Assembly Bill 2013 (AB 20213), Generative Artificial Intelligence: Training Data Transparency Act (2024), which came into effect on January 1, 2026. AB 2013 mandates that developers of generative AI systems released or substantially modified after January 1, 2022, and offered for public use in California, must publicly disclose a "high-level summary" of their training data on their official websites. The exemptions are systems for cybersecurity/security testing, aircraft operation in national airspace, national security/military/defense. AB 2013 does not establish opt-out rights for copyright holders, does not prohibit unlicensed training, and does not provide compensation mechanisms.
However, the viability of state-level AI regulation like AB 2013 is now in serious concern. On December 11, 2025, President Trump signed an Executive Order, Ensuring a National Policy Framework for Artificial Intelligence, which represents a dramatic reversal of regulatory momentum toward state autonomy. The Executive Order explicitly characterizes state AI laws as obstacles to U.S. global AI competitiveness, stating that they create 'a patchwork of 50 different regulatory regimes' that hinder innovation and U.S. dominance over China.

*China*

In China, the *Interim Measures for the Management of Generative Artificial Intelligence Services* were implemented on August 15, 2023. These regulations, comprising 24 articles, aim to strike a balance between fostering innovation and ensuring the security and governance of AI. Article 3 emphasizes the importance of maintaining a harmonious relationship between development and innovation while prioritizing security and governance in the field of AI. Articles 5 and 6 highlight the need for collaboration in developing basic technologies, such as chips and software platforms, as well as the creation of shared data resources. Article 16 states that all regulatory measures must be compatible with innovation, and Article 2 clarifies that the regulations apply only to publicly available generative AI services. Service providers are held responsible for the content generated using their services. Chinese regulations are among the most stringent in the world. For example, Article 12 mandates that users must be informed when content is generated using AI as a blanket rule.
On July 26, 2025, just tree days after Trump issued his AI Action Plan, China unveiled *Global AI Governance Action Plan*. The document details a comprehensive international framework for AI development and deployment. Unlike the U.S. plan, which centers on national priorities and addressing ideological biases in AI systems, China's approach focuses on global coordination, emphasizing multilateralism, openness, and technological support for developing countries. It does not generally address copyright protection in detail, but it does mention data and privacy protection, the lawful use of training data, and the need to actively explore regulated data transactions in AI. The document frames the creation of high-quality, legitimate data sets as essential for AI development, with safeguards in place, implying that intellectual property (including copyright) should be respected in the pursuit of innovative data governance.

*European Union*

On August 1, 2024, the European *Artificial Intelligence Act* (AI Act) entered into force (Regulation (EU) 2024/1689 of the European Parliament and of the Council of 13 June 2024, 2024). Article 53(1) of the EU AI Act, which became effective on August 2, 2025, establishes specific copyright-related obligations for providers of general-purpose AI (GPAI) models. Specifically, Articles 53(1)(c) and (d) require GPAI providers to: (i) implement a policy to identify and comply with reservation of rights expressed under the Copyright Directive (EU) 2019/790, and (ii) publish a sufficiently detailed summary of the content used for model training.

The Copyright Directive (EU) 2019/790 introduced two landmark exceptions for text and data mining that directly apply to AI training activities (Directive (EU) 2019/790 of the European Parliament and of the Council of 17 April 2019 on Copyright and Related Rights in the Digital Single Market and Amending Directives 96/9/EC and 2001/29/EC (Text with EEA Relevance.), 2019). These provisions establish a dual system: a broad exception with an opt-out mechanism for commercial use (art. 4), balanced against a narrower exception for scientific research (art. 3).

(*Commission Presents Template for General-Purpose AI Model Providers to Summarise the Data Used to Train Their Model*, 2025; *The General-Purpose AI Code of Practice*, 2025). The template balances transparency requirements with protection of intellectual property rights such as trade secrets, allowing providers to describe private datasets in a general manner without disclosing specific works. Importantly, the template supports the opt-out model, enabling rightsholders to reserve their rights against text and data mining through machine-readable protocols.

GPAI providers placed on the EU market before 2 August 2025 have until 2 August 2027 to publish their summaries in compliance with the new template. Importantly, the AI Act requires qualifying providers to comply with Article 53 regardless of where the GPAI model was trained.

To support implementation of the opt-out mechanism, the European Commission launched an official stakeholder consultation on copyright compliance under the EU AI Act and the GPAI Code of Practice, running from 1 December 2025 to 23 January 2026 (extended from the original 9 January 2026 deadline). The consultation, supported by the EU Intellectual Property Office (EUIPO), seeks to identify state-of-the-art, technically implementable machine-readable protocols for expressing reservations of rights against text and data mining (Uuk, 2025).

However, the implementation timeline of the EU AI Act has been subject to revision through *Digital Omnibus Regulation Proposal* (2025), unveiled by the European Commission on 19 November 2025. Most significantly for copyright compliance, Digital Omnibus does not postpone Article 53 copyright transparency requirements, which entered into force on 2 August 2025 and remain applicable. The Digital Omnibus proposal must still be approved by the European Parliament and Council, with adoption not expected before mid-2026.

The EU has been preparing a comprehensive strategy to accelerate the responsible and impactful use of artificial intelligence in science, with major policy developments expected by the end of 2025. One of the developments within this strategy was *Living guidelines on the responsible use of generative AI in research* (Directorate-General for Research and Innovation, 2025), as well as key initiatives such as a distributed AI infrastructure and a European AI Research Council (*Artificial Intelligence (AI) in Science*, 2025).
On October 8, 2025, the European Commission released its European Strategy for Artificial Intelligence in Science (A European Strategy for Artificial Intelligence in Science Paving the Way for the Resource for AI Science in Europe (RAISE), 2025), a document which promises to significantly influence copyright governance concerning AI training data within the scientific publishing ecosystem. The strategy explicitly acknowledges the necessity of gathering evidence to improve both access to and the reuse of publicly funded research outputs. This focus suggests a potential shift in policy, one that could broaden the permissible uses of openly accessible research for the purposes of training AI systems.
A central and pivotal element of the strategy is its emphasis on data governance, particularly through established initiatives like the European Open Science Cloud (EOSC) and the planned Data Labs within AI Factories. These structures imply a movement toward more formalized and structured permissions frameworks for data access. Yet, it is notable that the document remains largely silent on the specific copyright protections afforded to authors whose published works constitute the raw material for these data spaces. However, the strategy does not adequately differentiate between the long-accepted practice of text and data mining for non-commercial, human led research and the increasingly prevalent commercial application of AI training, a distinction with profound legal implications.
Whether the strategy's promise to update the "Living Guidelines on the responsible use of generative AI in research" and to establish new ethics review processes will provide a viable pathway for addressing these unresolved copyright concerns remains an open question. These proposed mechanisms may indeed offer a forum for such discussions, but the initial communication leaves considerable room for interpretation regarding the balance between fostering AI innovation and safeguarding intellectual property rights in academia.
The European Union's approach to AI regulation represents the most comprehensive and binding framework globally, with specific mechanisms to address copyright concerns that have been largely absent from other jurisdictions. While the EU AI Act initially focused primarily on risk management and transparency, the implementation through Article 53, the GPAI Code of Practice, and the ongoing stakeholder consultation process demonstrates a substantive effort to operationalize copyright protection in the AI training context. The opt-out model, requiring GPAI providers to respect machine-readable rights reservations, represents a significant advance over frameworks that merely defer to existing copyright laws. However, the Digital Omnibus proposal reveals ongoing political tensions between innovation promotion and regulatory enforcement, with industry pressure leading to delays in some areas while copyright compliance timelines remain intact. The ultimate effectiveness of these measures will depend on international adoption, technical standardization of opt-out protocols, and enforcement mechanisms yet to be fully developed.

### *Global Initiatives*

The OECD *AI Principles* (2019) set global, value-based guidance for trustworthy AI: promote inclusive growth and well-being; respect human rights, the rule of law and democratic values; ensure transparency and explainability; build robustness, security and safety; and maintain accountability across the AI lifecycle. Governments are urged to invest in R&D, enable high-quality and representative data, build interoperable ecosystems, adopt agile and outcome-based regulation (including experimentation), and prepare society and workers for AI-driven transformation. The Principles influence many national and international frameworks and emphasize privacy, non-discrimination, and responsible data use. At the same time, they do not create copyright rules or specifically resolve training-data copyright questions; rather, they root in existing IP and legal regimes and encourage accountability and transparency practices that can complement copyright compliance.

The UNESCO Recommendation on the Ethics of Artificial Intelligence (2022), adopted by all 193 member states in November 2021, establishes the first global standard for AI ethics grounded in human rights, dignity, and environmental protection. The framework promotes four foundational values (human rights and dignity; living in peaceful societies; ensuring diversity and inclusiveness; environment and ecosystem flourishing) and ten key principles including proportionality, human oversight, transparency, explainability, accountability, and fairness. It provides extensive policy action areas covering data governance, education, gender equality, culture, health, and environment, while requiring member states to implement ethical impact assessments and oversight mechanisms throughout AI lifecycles. While it calls for transparency about data sources and requires compliance with legal frameworks, the Recommendation does not establish specific copyright rules but emphasizes the importance of respecting existing intellectual property rights and international law.

The *G7 Hiroshima Process*, launched in May 2023 and finalized in December 2023, created international framework for advanced AI systems through the Hiroshima AI Process Comprehensive Policy Framework, which includes International Guiding Principles for all AI actors and a voluntary International Code of Conduct for organizations developing advanced AI systems (Hiroshima Process International Guiding Principles for Advanced AI System, 2023). However, the principles mention protecting intellectual property and implementing "appropriate data input measures and protection for personal data and IP" (Principle 11), but like other international frameworks, they defer to existing copyright laws rather than creating new IP protections.

**Table 1: AI-Related Regulations**

| Country | Document | Year | Main Points | Copyright Protection |
|---|---|---|---|---|
| United Kingdom | A Pro-Innovation Approach to AI Regulation | 2023 | Framework document establishing basic principles for AI; promotes flexible regulation; aims to foster innovation through AI development and use; no comprehensive AI regulation | No specific mention of copyright protection in the framework |
| United Kingdom | Artificial Intelligence (Regulation) Bill | 2025 | Private Members' Bill reintroduced by Lord Holmes; proposes creation of AI Authority for regulatory coordination; establishes regulatory sandboxes; requires designated AI officers; mandates independent AI audits; requires health warnings and labeling for AI products | Specifically requires organizations training AI to report all third-party data and IP used to AI Authority with assurances of informed consent and compliance with copyright laws; addresses transparency around training data provenance and IP usage |
| United Kingdom | Data (Use and Access) Act 2025 | 2025 | Does not immediately change copyright law. Establishes mandatory government assessment and reporting requirements (Sections 135-136). | Does not currently establish copyright protections but mandates comprehensive government assessment by March 18, 2026. |
| United States | Executive Order 14110 (RESCINDED) | 2023 | Safe, secure, and trustworthy development and use of AI; included actions related to copyright law; recommendations on copyright and AI issues | Specifically addressed copyright issues, including scope of protection for AI-produced works and treatment of copyrighted works in AI training |
| United States | Generative AI Copyright Disclosure Act (H.R.7913) | 2024 | Requires companies to notify U.S. Copyright Office about copyrighted works used in AI systems 30 days before/after public release; ensures transparency and accountability | Directly addresses copyright by requiring disclosure of copyrighted training data, but criticized for being vague and biased toward AI developers rather than copyright holders |
| United States | Executive Order 14179 & AI Action Plan 2025 | 2025 | Removes barriers to "American AI leadership"; prioritizes "American values" like free speech; supports open models; streamlines infrastructure permitting; highly politicized approach | No mentions of copyright found in the document, representing a departure from previous copyright considerations |
| United States (California) | Assembly Bill 2013 (AB 2013): Generative AI Training Data Transparency Act | 2025 | Applies to developers of generative AI systems released or substantially modified after Jan. 1, 2022, made available to Californians (free or paid). Developers must post on their website a "high-level summary" of training data. | First U.S. legislation specifically targeting copyright in AI training data disclosure. Requires public disclosure of intellectual property status of training datasets, including copyright, trademark, and patent status. The law does not provide an opt-out mechanism. There is a serious concern that this law will be blocked by the federal government. |
| China | Interim Measures for the Management of Generative AI Services | 2023 | 24 articles balancing innovation with security/governance; applies only to publicly available services; service providers responsible for generated content; mandatory AI labeling | Does not specifically address copyright protection; focuses on content control and governance rather than IP rights |
| China | Global AI Governance Action Plan 2025 | 2025 | Comprehensive international framework emphasizing multilateralism, openness, and technological support for developing countries; focuses on global coordination | Mentions data and privacy protection, lawful use of training data, and need for legitimate datasets, implying IP rights should be respected but no specific copyright provisions |

| European Union | Copyright Directive (EU) 2019/790 (Digital Single Market Directive) | 2019 | Introduces two text and data mining (TDM) exceptions: Article 3 (mandatory exception for scientific research with no opt-out mechanism) and Article 4 (broad exception for all TDM including commercial use, with opt-out mechanism); rightsholders can reserve rights through machine-readable means for publicly available online content | Establishes opt-out framework for AI training: Article 4(3) allows rightsholders to expressly reserve rights against TDM in “appropriate manner” (robots.txt, metadata, terms of service for online content); default permits TDM without authorization unless opt-out expressed; directly linked to EU AI Act Article 53(1)(c) which requires GPAI providers to identify and respect these reservations; Article 3 provides mandatory exception for scientific research without opt-out option |
|---|---|---|---|---|
| European Union | EU Artificial Intelligence Act (AI Act) | 2024 | Entered force August 1, 2024; reduces AI-associated risks; focuses on high-risk AI systems; requires transparency for low-risk systems; mandatory labeling of AI-generated content | Articles 53(1)(c) and (d) require GPAI providers to:<br>- implement a policy to identify and comply with reservation of rights<br>- publish a “sufficiently detailed summary” of the content used for model training |
| European Union | GPAI Code of Practice | 2025 | Voluntary code providing detailed implementation guidance for EU AI Act obligations; three chapters on transparency, copyright, and safety; commits signatories to respect machine-readable opt-out protocols; requires copyright policies and safeguards against infringing outputs | Directly addresses copyright through five concrete measures: copyright policy requirement; lawful data access; respect for rights reservations ("do not train" signals); mitigation of infringing outputs; complaint mechanisms for rightsholders |
| European Union | Article 53 Training Content Summary Template | 2025 | Template published by AI Office requiring GPAI providers to disclose websites from which they sourced training data; balances transparency with trade secret protection; applies to models placed on EU market before 2 August 2025 with compliance deadline of 2 August 2027 | Implements opt-out model, i.e. enables rightsholders to reserve rights against text and data mining; requires disclosure of training data sources but protects proprietary information; extraterritorial effect regardless of where model was trained |

| European Union | EU Commission Consultation on TDM Opt-Out Protocols | 2025-2026 | Stakeholder consultation (1 December 2025 - 23 January 2026) to identify machine-readable protocols for expressing rights reservations; supported by EUIPO; seeks agreement on common list of opt-out solutions for GPAI Code signatories | Operationalizes copyright opt-out mechanism; identifies technical standards for rightsholders to reserve rights under Copyright Directive Article 4(3); establishes regularly reviewed list of agreed machine-readable solutions |
|---|---|---|---|---|
| European Union | Digital Omnibus Proposal | 2025 | Proposes to delay enforcement of some provisions from August 2026 to December 2027; response to industry lobbying and concerns about implementation readiness; does not delay Article 53 copyright requirements | Does not affect copyright compliance timeline: Article 53 transparency and opt-out requirements remain applicable on original schedule |
| European Union | Living Guidelines on Responsible Use of Generative AI in Research | 2025 | Part of comprehensive strategy to accelerate responsible AI use in science; aims to enhance innovation, competitiveness, ethical deployment, and international leadership | Does not explicitly address copyright protection; mentions only "unpublished and sensitive work" but no specific IP protections |
| International (OECD) | OECD AI Principles | 2019 | Global, value-based guidance for trustworthy AI; promotes inclusive growth, human rights respect, transparency, robustness, and accountability; influences many national frameworks | Does not create specific copyright rules; defers to existing IP and legal regimes; encourages accountability and transparency that can complement copyright compliance |
| International (UNESCO) | AI Ethics Recommendation | 2021 | Adopted by 193 member states; first global standard for AI ethics; four foundational values and ten key principles; extensive policy action areas | Does not establish specific copyright rules but emphasizes respecting existing intellectual property rights and international law; requires compliance with legal frameworks |
| International (G7) | Hiroshima Process Comprehensive Policy Framework | 2023 | International framework for advanced AI systems; includes International Guiding Principles and voluntary Code of Conduct; emphasizes safe, secure, trustworthy AI development | Mentions protecting intellectual property and implementing "appropriate data input measures and protection for personal data and IP" but defers to existing copyright laws rather than creating new IP protections |

## Copyright, Licensing, and Legal Analysis

### *Copyright Framework and Creative Commons Licensing*

Most scientific works are protected by copyright laws. Copying and retaining these works in AI systems, as well as reproducing them in outputs, involves copyright, making appropriate licensing essential for compliance (Johnson, 2024). The generated output can be considered a derivative work, although this is not explicitly stated in any legal documents.

Creative Commons (CC) licenses are the most widely used for open-access outputs. Approximately 28% of global research output is licensed under the Creative Commons Attribution license (CC BY), while another 22% uses more restrictive Creative Commons licenses (Pollock & Michael, 2024). However, Creative Commons acknowledges that existing CC licenses do not fully address the specific challenges related to using creative works for AI training (Walsh, 2023).

On the other hand, if the use of content is subject to copyright exclusions, the licensee's abilities are limited. In fact, such an exclusion is currently being considered for legislation in the United States. Moreover, the U.S. fair use doctrine allows for the unlicensed use of copyrighted works under certain circumstances. AI training is often considered a case of fair use (Johnson, 2024; Walsh, 2023). For instance, OpenAI argues that this position is "supported by long-standing and widely accepted precedents" (*OpenAI and Journalism*, 2024).

Publishers are also responding to market changes by developing licensing agreements for the use of content in LLM training (Schonfeld, 2024). Currently, the number of such deals is relatively low[2], and they primarily cover content distributed through subscription services. If a publishing contract includes the full transfer of rights to the publisher, the publisher can license the content for AI training without seeking the authors' consent (Hansen, 2024). This underscores the importance of the rights retention strategy. Major publishers, along with Clarivate, are rapidly developing new AI-based businesses, which are evolving into data cartels (Pooley, 2024). This could lead to a situation where the academic AI market adopts the same oligopolistic structure as the current academic publishing market.

In June 2025, Creative Commons unveiled CC Signals ("Introducing CC Signals," 2025), an initiative widely regarded as the most substantial development in open licensing since the suite's initial introduction. This move responds to a recognized shortfall in the capacity of existing Creative Commons licenses to govern rights concerning AI training data. Rather than operating as a simple binary of permission or denial, CC Signals proposes a system of preference signals. Its stated aim is to increase reciprocity and "sustain the commons in the age of AI." This will allow dataset curators to express conditional preferences for machine-based content reuse. A critical aspect of the design is its dual nature as both a technical and a legal instrument. The signals are engineered to be machine and human readable, a feature intended to provide flexible application across what might be termed legal, technical, and normative contexts. One might ask, however, whether such flexibility can be standardized effectively across diverse jurisdictions.

An alpha version was initially planned for release in November 2025; however, as of January 2026, CC Signals remains in development phase. The development process is being conducted, in the words of Creative Commons, "alongside our partners and community," a commitment evidenced by the solicitation of input through a GitHub repository (*Creativecommons/Cc-Signals*, 2025/2025). This approach suggests an iterative methodology where the technical specifications are being refined in direct response to stakeholder feedback, a process that may prove crucial for the framework's eventual adoption and legitimacy. The ultimate success of this endeavor, of course, remains to be seen, depending on complex factors of both technical implementation and community buy-in.
The Really Simple Licensing (RSL) standard, launched September 2025 and formalized as RSL 1.0 in December 2025 (Eayrs, 2025), represents the next step and a potentially significant moment in machine-readable licensing frameworks. RSL operates as an open standard enabling publishers to embed machine-readable licensing terms directly into web content metadata. Unlike static legal documents, RSL terms execute automatically when AI crawlers access content, eliminating the need for bilateral negotiation or centralized licensing intermediaries. RSL embeds licensing preferences directly into robots.txt files, HTTP headers, RSS feeds, and HTML metadata. Possible compensation models include (Eayrs, 2025):

1. *Attribution-Only License:* free machine access provided visible credit and functional links to original sources are included in AI outputs.
2. *Pay-Per-Crawl:* automated micropayments triggered each time an AI system accesses copyrighted content.
3. *Pay-Per Inference:* Developers pay creators a micro-royalty for each inference (output generation) their AI model produces.
4. *Attribution + Reciprocity:* AI developers commit to supporting content ecosystems through donations to non-profits, shared datasets, or open-source model releases.

[2] Generative AI Licensing Agreement Tracker. URL: https://sr.ithaka.org/our-work/generative-ai-licensing-agreement-tracker/.

Right holders also have an option to prohibit AI use and search summaries (*Really Simple Licensing (RSL) 1.0 Specification*, 2025).
CC Signals and RSL standard shift the narrative from (non)transformative and (non)commercial use to the question “What did the publisher declare acceptable?”. However, RSL compliance remains entirely voluntary with no binding legal requirement for AI companies to honor machine-readable licensing terms. Without partnerships with content delivery networks like Fastly or Cloudflare to implement technical barriers blocking non-compliant crawlers, publishers can request payment but lack mechanisms to enforce it.
A further complication involves the tricky question of retroactive application. The scholarly ecosystem already contains millions of research outputs shared under existing CC BY licenses, which cannot be easily relicensed. It is therefore questioned whether CC Signals will apply solely to new publications. Such a limitation would potentially create a two-tiered system, a situation where a vast corpus of older research remains freely trainable by AI systems without any reciprocity requirements, thereby diluting the framework's overall impact.

*Legal Analysis of AI-Generated Content as a Fair Use Case*

The fair use analysis applied to AI training practices has yielded deeply divided judicial opinions. This legal uncertainty is extensively documented in comparative legal scholarship (Chopra et al., 2025; Sag & Yu, 2024), which highlights how different legal traditions produce varied outcomes in analogous cases. Recent litigation highlights the critical importance of two factors: the manner in which training data is sourced and the commercial nature of the eventual AI application.
A persistent and legally uncertain question is whether content produced by AI systems should be classified as a derivative work of the copyrighted materials utilized during the training process. Courts have thus far provided inconsistent guidance on this matter. Under the framework of the U.S. copyright law, a derivative work is legally defined as one “based upon one or more preexisting works,” which constitutes a “recasting, transformation, or adaptation” of an original source (*17 U.S. Code § 101 - Definitions*, n.d.).
In May 2025, the U.S. Copyright Office released Part 3 of its comprehensive Report on Copyright and Artificial Intelligence, directly addressing whether the training of generative AI systems on copyrighted works constitutes fair use (*Copyright and Artificial Intelligence, Part 3: Generative AI Training*, 2025). The main conclusion is that AI training does not automatically qualify as a fair use case, each case requires thorough analysis under the four statutory fair use factors – purpose and character, nature of the copyrighted work, amount and substantiality, and market effect.

Proponents of classifying AI output as derivative often argue that these outputs can, in certain circumstances, incorporate protectable expression from the training dataset. This is particularly plausible when the generated content bears a close resemblance to specific, identifiable source materials. At the same time, the degree of similarity substantial enough to consider the work a derivative is not entirely clear (Griem, Jr. & Wallace, 2023).

Conversely, a strong counterargument suggests that AI models do not store or replicate copies but instead learn to generate content based on abstract statistical patterns. From this perspective, the outputs represent novel combinations that do not constitute a direct adaptation of any specific pre-existing work. This view found judicial support in *Kadrey v. Meta Platforms, Inc., No. 3:24-cv-02029* (2025), where the Northern District of California dismissed the claim that an LLM is itself an infringing derivative work. The court deemed this notion "nonsensical," reasoning that there is no way to understand an LLM as a recasting or adaptation of the plaintiffs' books. The court also rejected the broader proposition that every output from an LLM is automatically a derivative work, insisting instead on a case-specific demonstration of substantial similarity.

The settlement in *Bartz v. Anthropic PBC, No. 3:24-cv-05417* (2025) points toward a more nuanced legal pathway. In that case, Judge Alsup determined that training the Claude model on legally acquired books was a "transformative" fair use. Nevertheless, the subsequent $1.5 billion settlement represents a watershed moment in this area of litigation. The lead plaintiff, Andrea Bartz, characterized the $1.5 billion settlement as a clear message: "You are not above the law, our intellectual property isn't yours for the taking" (Ortutay, 2025). The settlement draws a sharp distinction based on data provenance. While training on legally obtained books was deemed potentially fair, Judge Alsup ruled that Anthropic's use of "pirated" copies from shadow libraries was inherently infringing."

Building on the momentum of the *Bartz v. Anthropic* class action settlement and explicitly rejecting its purportedly "bargain-basement" terms, Pulitzer Prize-winning journalist John Carreyrou filed individual copyright infringement actions on December 22, 2025, against six major AI companies on behalf of himself and five fellow authors (*Carreyrou v. Anthropic PBC et al., 25-cv-10897*, 2025). This lawsuit signals that the copyright community views the Anthropic settlement not as a resolution but as an inadequate interim measure, and that subsequent litigation will likely demand accountability for the full scope and value of training data misappropriation, potentially triggering damages awards that would fundamentally reshape AI companies' business models and their obligations to compensate creative rights holders.

Arguments against fair use, however, concentrate on commercial harm and the potential for market substitution. A pivotal moment came from the Delaware District Court in *Thomson Reuters Enterprise Centre GmbH v. Ross Intelligence Inc., No. 1:20-cv-613-SB* (2025), which issued the first unambiguous rejection of a fair use defense for AI training. The court concluded the use was "commercial" and, importantly, "not transformative." A key element of the ruling was the recognition of an "obvious" potential market for licensing copyrighted works specifically for AI training. The court took judicial notice that more and more copyright owners are striking deals with AI companies to license works for training purposes, thereby substantiating the market harm factor.

A critical evidentiary ruling issued on January 5, 2026, in the consolidated copyright litigation against OpenAI (*In Re: OpenAI, Inc. Copyright Infringement Litigation*, 2026) directly undermines OpenAI's fair use defense by requiring production of 20 million de-identified ChatGPT user interaction logs. The District Court affirmed Magistrate Judge Ona T. Wang's

November 7, 2025 discovery order, rejecting OpenAI's privacy and burden objections and holding that the full unfiltered log sample is essential because even logs which do not reproduce plaintiffs works may help OpenAI assert defenses such as fair use and are thus relevant for this case under the applicable discovery standard.

Europe demonstrates completely different legal patterns. The Munich Regional Court issued a comprehensive decision on November 11, 2025, in GEMA v. OpenAI (Case No. 42 O 14139/24) finding that both the training of ChatGPT using protected song lyrics and the subsequent output of those lyrics constitute copyright infringement under German law (*Entscheidung GEMA v. OpenAI, 42 O 14139/24*, 2025). The court rejected OpenAI's reliance on the EU text and data mining (TDM) exception under Directive 2019/790, holding that the reproduction of song lyrics within model parameters exceeds the scope of permissible TDM, which is limited to extraction of information such as abstract syntactic regulations, common terms and semantic relationships rather than "memorization" of entire creative works. Critically, the court placed full responsibility on OpenAI as the developer and operator of the system, rejecting arguments that user prompts create independent liability and concluding that OpenAI directly commits the act of communicating to the public by designing and operating a system capable of reproducing protected works.

Beyond the training-data copyright disputes exemplified by *Bartz v. Anthropic* and *Kadrey v. Meta*, a separate class of AI copyright infringement has emerged involving retrieval-augmented generation (RAG) systems that reproduce copyrighted content in real-time response to user queries, rather than through initial model training. On December 5, 2025, The New York Times Company filed a comprehensive copyright and trademark infringement complaint against Perplexity AI, Inc. (*The New York Times Company v. Perplexity AI, Inc., Complaint for Copyright Infringement, Trademark Infringement, and Related Relief, Case No. 1:25-cv-10106*, 2025), alleging that Perplexity's AI-powered search engine systematically copies and reproduces Times' content without authorization, despite 18 months of cease-and-desist letter.

The significance of New York Times v. Perplexity lies in its articulation of a two-stage infringement model that challenges the assumption that fair use protections extend to all forms of AI content processing. In *Stage 1 (Input)*, the complaint alleges that Perplexity deployed specialized web crawlers (PerplexityBot and Perplexity-User-Agent) that made over 175,000 attempts to access nytimes.com in August 2025 alone, deliberately circumventing The Times' robots.txt protocol and hard-block technical barriers designed specifically to prevent scraping. The complaint characterizes this conduct as willful, since The Times had previously notified Perplexity by cease-and-desist letters to cease accessing its content, the continued scraping after express objection establishes the requisite intent for enhanced damages.

In *Stage 2 (Output)*, the complaint alleges that Perplexity's GenAI products generate responses that are "identical or substantially similar" to Times content, reproducing entire articles or substantial portions thereof in "verbatim or near-verbatim" forms that provide users with comprehensive answers eliminating any need to visit the source material. The Times provides specific examples.

The complaint asserts that this two-stage process constitutes infringement that cannot be shielded by fair use. The key distinction from prior AI training cases is that fair use analysis has typically focused on transformative creation of new works from training data. By contrast, Perplexity's RAG system creates no new work; it retrieves existing copyrighted content and reproduces it, adding only minor synthesizing commentary.

Table 2 summarizes the analyzed legal cases in terms of fair use statutory factors.

**Table 2: Four-Factor Fair Use Test Application in the AI Copyright Cases (2025)**

| Case and Court | Factor 1: Purpose and Character of Use | Factor 2: Nature of the | Factor 3: Amount and | Factor 4: Market Effect | Fair Use Outcome |
|---|---|---|---|---|---|

| | | Copyrighted Work | Substantiality Used | | |
|---|---|---|---|---|---|
| Kadrey v. Meta Platforms, Inc. | *Transformative:* commercial AI training creates fundamentally new tool (general-purpose LLM) distinct from original novels | *Favors fair use:* novels are creative works, but transformative use of training data mitigates | *Disfavors fair use:* entire books used as training data; substantial copying | *Favors fair use:* no market harm demonstrated; plaintiffs failed to prove AI outputs competed with book sales; no licensing market shown | *Fair use upheld:* training on lawfully acquired books constitutes fair use despite commercial nature |
| Thomson Reuters Enterprise Centre GmbH v. Ross Intelligence Inc. | *Not transformative:* commercial AI tool directly competes with Westlaw | *Disfavors fair use:* copyrighted summaries and original documents | *Disfavors fair use:* entire copyrighted headnotes directly copied into training data; substantial amount of identifiable works reproduced | *Strong market harm:* direct competition with Westlaw legal research platform; emerging licensing market for legal data to train AI | Fair use rejected: partial summary judgment for Thomson Reuters; court found non-transformative commercial infringement with clear market substitution |
| Bartz v. Anthropic PBC | *Mixed:* training on lawfully acquired books = *transformative*; training on pirated books = *not transformative*; commercial model | *Disfavors fair use:* books (creative works); full copyright protection | *Disfavors fair use:* millions of entire books used; critical distinction drawn between lawful and pirated sources | *Significant market harm:* market harm from piracy; emerging licensing deals with publishers; settlement value ($1.5B) acknowledges infringement liability | *Settled for $1.5 billion* Partial fair use for lawful training; pirated book training admitted infringing; data provenance (lawful vs. pirated) is a decisive factor |
| The New York Times Company v. Perplexity AI, Inc. *(compliant filed – early-stage litigation!)* | *Not transformative:* real-time content retrieval and reproduction to end users; direct market substitute for visiting nytimes.com; RAG system retrieves and reproduces, not creates | *Disfavors fair use:* original journalism; creative editorial composition and analysis; favors copyright holder | *Disfavors fair use:* verbatim or near-verbatim reproduction of entire articles; specific examples include complete article copying | *Perfect market substitution:* direct competition with nytimes.com; users obtain comprehensive content summaries without visiting the source; lost pageviews, ad revenue, subscriptions | *Alleged infringement* Complaint alleges two-stage infringement (scraping + output reproduction); willful infringement alleged |
| GEMA v. OpenAI | *Not transformative:* paid commercial | *Disfavors fair use:* song lyrics (creative musical | *Disfavors fair use:* entire song lyrics | *Direct market harm:* eliminates | *Fair use rejected:* |

| (German law) | subscription model; memorization of protected song lyrics for commercial AI product | pieces); full copyright protection under German law | embedded in model parameters; memorization creates retrievable reproductions matching source lyrics | licensing opportunity; copyright holders unable to control reproductive uses | German court found TDM exception inapplicable; memorization exceeds permissible use; ordered cease-and-desist + damages; rejects EU TDM exception for AI training |
|---|---|---|---|---|---|

## A Call for Action

Science and artificial intelligence (AI) are closely linked. Research provides data, which is crucial for training large language models (LLMs) and advancing data science more broadly. At the same time, generative AI (GenAI) is revolutionizing research. Open-source LLMs are an essential part of open science. While AI presents significant opportunities for scientific advancement, it also poses substantial risks. Legislation in this field is still evolving, and regulatory and policy documents often focus on attracting investment in AI or promoting its responsible development and use. The use of publicly available research outputs for training LLMs falls into a "grey area." At the moment, the community lacks any meaningful discussion on the reuse of academic content for LLMs' training. Attempts to raise this issue are made, but their impact is rather limited (Decker, 2025). Below, I offer some thoughts on actions that can be taken in the near future.

First and foremost, AI training should not be considered an exception to copyright law (i.e., under the fair use doctrine). Recognizing LLM training as a case of fair use undermines efforts to reform copyright regulation. In my opinion, LLM training should not qualify as fair use for at least three main reasons:

1. Many AI systems already operate on paid subscription models. Even if no fees are currently charged, there are no legal restrictions preventing these models from becoming commercialized in the future.
2. AI-generated content often closely resembles the original, making it subject to copyright and attribution requirements. It's a derivative work (not transformative)!
3. AI-generated output, not referencing the source, violates academic traditions of credit allocation. It can be considered a special case of "market harm."

This issue is particularly relevant in the U.S. context but given that most AI developers are based in the United States, it is critical for the global development of the industry. Some researchers argue that it will take years for U.S. courts to address the issue of licensing content for LLM training (Bergstrom & Ruediger, 2024). This is a major concern for the academic community, as the market will continue to evolve, researchers will increasingly rely on AI for interacting with scholarly output, and it will become more difficult to implement changes (see below for further discussion of limitations and challenges).

Authors should have the option to refuse the use of their work for training GenAI models or specific groups of such models. On the one hand, in terms of Creative Commons licensing, there are two possible strategies to achieve this:

1. *Examine existing licenses:* The Creative Commons BY-ND (Attribution-NoDerivatives) license could be considered restrictive for AI training, but only if regulatory frameworks recognize AI-generated content as derivative works. However, determining whether AI-generated content qualifies as a derivative work is complicated by the fact that LLMs can produce different responses for each query, making it difficult to assess similarity to the original. The BY-NC (Attribution-NonCommercial) license may also be restrictive for training models intended for commercial use[3].

2. *Introduce a new "NT" (no train) extension:* This would explicitly prohibit the use of licensed works for AI training. However, since the original datasets used for LLM training are not publicly accessible, the prospects for enforcing such licensing terms remain uncertain. Additionally, publishing contracts should specify that publishers cannot use articles to train their LLMs or other AI models without author consent.

On the other hand, the evident limitations of existing Creative Commons licenses in governing AI training practices, as previously discussed, appear to have motivated the development of CC Signals. This new framework for expressing preferences is conceived specifically for contemporary AI applications. Instead of depending on interpretations of legal tools designed for human-centric content sharing, CC Signals proposes purpose-built mechanisms to articulate permissions and constraints for machine learning processes.

From a personal perspective, there is a compelling case for the academic community to rally behind the CC Signals initiative, rather than pursuing an isolated NT license extension. The CC Signals framework, despite its ongoing development, seems to offer a more comprehensive approach. It attempts to grapple with the intertwined challenges of enforceability, machine readability, and the crucial need for international coordination. The advantage of CC Signals lies in its institutional backing from a trusted organization, its participatory development process, and its explicit design for systems-level coordination across multiple domains.

Consequently, universities and major research funders would be well-advised to commit to adopting CC Signals, though such commitment should be contingent on the framework meeting specific requirements. These would include the provision of legally binding signals where jurisdictionally feasible, a default position that opts-in to commercial AI training unless otherwise specified, clearly articulated definitions of what constitutes adequate reciprocity, and the development of mechanisms for retroactive application to content already published under traditional CC licenses, though this presents obvious practical difficulties.

Another emerging mechanism is the Really Simple Licensing (RSL) standard, formalized as RSL 1.0 in December 2025. Unlike static copyright licenses, RSL terms execute automatically when AI crawlers access content, because the licensing preference is embedded directly into robots.txt files, HTTP headers, RSS feeds, and HTML metadata.

### *Universities as Key Players in AI Regulation*

Universities should take a leading role in regulating AI. On the one hand, universities often act as publishers or maintain their own repositories, making it feasible to implement content licensing approaches in practice. On the other hand, universities conduct research and develop GenAI models, placing them at the forefront of addressing the ethical aspects of these processes. Furthermore, universities can provide evidence to support legislative regulation. Having said that, I must acknowledge that universities lack the regulatory power that governments possess. Among specific actions that universities can take I would mention:

- Supporting development and implementation of CC Signals and RSL,
- Adopting institutional policies that prohibit the use of faculty work for AI training without consent,
- Developing open-source, responsible LLMs,
- Creating mandatory AI usage and AI ethics curricula.

[3] However, can we be certain that today's open models will not be commercialized in the future?

Most community documents in open science remain silent on the matter of AI training, e.g. the recent Barcelona Declaration on Open Research Information (Kramer et al., 2024). This apparent oversight represents a significant missed opportunity for the open science community to establish a coherent normative position on this issue. Such a position is arguably needed before commercial practices and legal expectations become entrenched and more resistant to change. The current moment therefore presents a critical juncture. It seems imperative that the community acts now to shape governance frameworks, ensuring that the principles of open science ultimately serve broader human flourishing rather than facilitating primarily corporate extraction.

### *Legislative Measures and International Cooperation*

Governments and international organizations must develop and implement legislative measures to protect authors' rights and prevent unauthorized use of their works for training GenAI models. One of the first steps should be the mandatory disclosure of training datasets by developers.

The challenge lies not only in adopting national AI laws but also in harmonizing these laws globally. Without international coordination, commercial developers could exploit "safe harbors" to serve their own interests. Therefore, it is essential for large intergovernmental organizations, such as UNESCO, to take on this task. International copyright harmonization for AI training would require establishing minimum standards while allowing national flexibility, similar to the TRIPS Agreement model (*Agreement on Trade-Related Aspects of Intellectual Property Rights*, 1994). Such harmonization requires at least three components: law alignment, procedural standardization, and enforcement coordination ("Harmonization of International Copyright Standards," 2025).

Another challenge is that AI models cannot be "untrained." If restrictions are imposed only on new models, existing models would gain a non-market advantage. Conversely, applying restrictions retroactively to existing models could destabilize the industry. A responsible dialogue is needed to find a balanced solution. One possible approach is retrieval-augmented generation, which allows models to reference relevant papers in their outputs ("AI Firms Must Play Fair When They Use Academic Data in Training," 2024).

## Conclusion

The author of this article does not oppose AI. In fact, while writing this manuscript, I used Yandex.Translate (YaGPT-5) and DeepSeek R3 to assist with reading Chinese and German source texts and proofreading the final version the paper, as well as Perplexity to search for relevant sources. This analysis argues that existing copyright frameworks are fundamentally ill-equipped to regulate the use of scholarly works in AI training, thereby posing urgent challenges to academic integrity and the equitable production of knowledge. Three principal findings arise from this investigation.

Widely implemented licensing mechanisms such as Creative Commons licenses prove inadequate for addressing the distinct challenges presented by AI training. Although CC BY licenses function effectively for human-to-human content sharing, they were never intended to govern machine learning processes at scale, nor do they adequately differentiate between non-commercial research applications and commercial AI development. The 2025 launch of CC Signals constitutes a significant institutional reaction to this problem. In December 2025, the RSL Technical Steering Committee published a machine-readable licensing standard RSL 1.0., enabling different models of compensation and direct prohibition. Nevertheless, potential efficacy of these advances depends on resolving fundamental questions concerning enforceability, its retroactive application, and the necessity for international coordination.

### *Fragmented Regulation and Judicial Practice*

Secondly, findings reveal fundamentally divergent regulatory approaches among major jurisdictions in addressing copyright protection for AI training data. While the United States initially attempted to address copyright issues under the Biden administration through Executive Order 14110 and the proposed Generative AI Copyright Disclosure Act, the Trump administration's 2025 AI Action Plan marks a significant shift in priorities, moving away from the copyright-focused approach of the previous administration. This creates a regulatory environment that critics argue disproportionately favors the interests of AI developers over those of copyright holders. However, California's Assembly Bill 2013 (AB 2013), which took effect on January 1, 2026, represents a notable state-level countermovement, establishing mandatory disclosure requirements for the intellectual property status of AI training datasets. Yet this federal-state regulatory fragmentation creates additional uncertainty: while AB 2013 requires transparency on copyright status, the Trump Administration's December 2025 Executive Order simultaneously attempts to block state AI laws through federal litigation.
The European Union has emerged as the global leader in establishing enforceable copyright protections for AI training. Building upon the Copyright Directive (EU) 2019/790, which established a text and data mining opt-out framework under Article 4(3), the EU AI Act's Article 53 creates binding obligations for general-purpose AI providers to identify and respect rightsholders' reservations of rights, publish detailed training content summaries, and implement policies to prevent copyright-infringing outputs. The GPAI Code of Practice, adopted in July 2025, translates these requirements into five concrete measures including respecting machine-readable opt-out protocols and establishing complaint mechanisms for rightsholders. Critically, the European Commission's ongoing stakeholder consultation (concluding January 23, 2026) aims to standardize machine-readable opt-out protocols, addressing a key implementation challenge. Unlike other jurisdictions that merely recommend or defer to existing copyright laws, the EU has created an operationalized opt-out model with extraterritorial reach. Significantly, while the Digital Omnibus proposal delays certain high-risk AI obligations until 2027 in response to industry lobbying, Article 53 copyright transparency requirements that entered into force on August 2, 2025, remain unaffected by these delays, demonstrating the EU's commitment to prioritizing intellectual property protection even amid pressure to ease regulatory burdens.
The UK's 2025 Artificial Intelligence (Regulation) Bill represents another significant development, directly addressing the copyright protection gap by requiring organizations training AI systems to report all third-party data and intellectual property used, with assurances of informed consent and compliance with copyright laws. The Data (Use and Access) Act 2025, which received Royal Assent on June 11, 2025, establishes a statutory framework mandating the government to publish by March 18, 2026 both an economic impact assessment and a comprehensive report on the use of copyright works in AI development, addressing three competing policy objectives: control, access, and transparency. International frameworks from OECD, UNESCO, and G7 continue to defer to existing copyright laws rather than establishing AI-specific IP protections, while China's regulatory measures remain focused on risk management and content control rather than addressing fundamental copyright issues in AI training.
The trajectory of 2025 fair use cases reveal system tensions. U.S. courts oscillate between protecting AI innovation (Kadrey v. Meta) and recognizing legitimate copyright claims (Thomson Reuters v. Ross Intelligence; New York Times v. Perplexity AI (filed)). International jurisdictions have moved decisively toward copyright enforcement: the Munich Regional Court rejected fair use defenses in GEMA v. OpenAI, finding that memorization of song lyrics in commercial LLMs exceeded permissible text-and-data-mining uses under German law and EU Copyright Directive. This conceptual instability, reflected in the U.S. split between transformativeness-focused defenses (Kadrey) and market-harm-based liability (Thomson

Reuters, Perplexity), has catalyzed two alternative frameworks that may reshape AI content governance without awaiting further judicial consensus or international harmonization.

*Towards Implementation*

These findings collectively support the idea of harmonizing international legislative efforts. The objective of such efforts would be to ensure transparency, protect intellectual property, and prevent the emergence of an oligopolistic market structure that could potentially prioritize commercial profit over research integrity and equitable knowledge production.

Achieving this harmonization demands three interconnected components. First, substantive law alignment is needed to establish minimum standards for author consent and disclosure, while still permitting national flexibility in implementation mechanisms, perhaps following models like the TRIPS Agreement. Second, procedural standardization should focus on creating machine-readable permission frameworks, CC Signals and RSL standard are examples, which would enable automated compliance verification at a large scale. Third, enforcement coordination is essential to prevent regulatory arbitrage, a situation where AI companies might exploit jurisdictional differences by training models in permissive legal environments for subsequent global deployment.

*Fundamental Choice*

The academic community stands at a critical juncture. Current market dynamics disproportionately benefit large AI developers, who can train models on copyrighted scholarly works without providing compensation or attribution. This pattern risks replicating the very oligopolistic concentration that already plagues the academic publishing industry. Therefore, universities, research funders, and scholarly societies must act decisively to influence governance frameworks before these commercial practices become entrenched. Specific actions could include the institutional adoption of CC Signals and RSL standard, contingent upon their binding enforceability; the development of new rights retention contract that explicitly prohibits unauthorized AI training; and coordinated advocacy for legislative reforms that establish firm transparency and consent requirements.

The central question, then, is not whether AI should be utilized in research, but rather whether its development will proceed in a manner that respects the intellectual property rights of knowledge creators, or if it will perpetuate extractive models that ultimately undermine the foundations of scholarly communication. This analysis suggests that without coordinated action across stakeholder groups and jurisdictions, the latter outcome appears increasingly inevitable. The time for incremental measures has passed; what is needed now is the immediate establishment of comprehensive governance frameworks.

It may appear that I oppose reuse, which is a fundamental part of the open science agenda. However, my critique targets specifically the commercial exploitation of research outputs absent appropriate attribution and remuneration. I argue that establishing transparent governance frameworks for industry actors, coupled with explicit rights reservation mechanisms for authors, will ultimately strengthen and advance open science sustainability.

*Limitations of the Study*

This analysis is subject to several important limitations that affect its scope and temporal applicability. In terms of geographic coverage, the study focuses on major economies and global initiatives but does not sufficiently engage with significant regional actors (South Korea, Taiwan, Israel). These regions may have developed distinctive regulatory or judicial approaches that could enrich the global understanding of AI copyright issues.

The further constraint is temporal. The field of AI and its corresponding regulatory environment are in a state of constant and rapid change. New legal precedents, settlement agreements, and regulatory frameworks are continuously emerging. Consequently, the information presented here is most accurate as of its publication date and may require revision as the legal landscape evolves.

The jurisdictional focus of the legal analysis is predominantly on U.S. copyright law and recent American court decisions. There is limited coverage of how analogous issues are being regulated in other major legal systems, such as the European Union, China, or the United Kingdom, or in emerging economies with growing AI sectors where different legal traditions may yield different answers.

Finally, a call for action mainly addresses open science outputs and relevant licensing. This area represents a major gap in regulation – copyright infringements are covered much better.

### *Future Research Directions*

This analysis reveals several promising avenues for further research:

1. Systematic examination of how different legal systems approach AI copyright issues, particularly in civil law jurisdictions and countries with different fair dealing/fair use doctrines.
2. Empirical research on the actual economic effects of AI training on copyright holders, including quantification of market harm and analysis of emerging licensing markets.
3. Interdisciplinary research examining how specific AI architectures and training methodologies affect legal analysis, particularly regarding substantial similarity and transformative use determinations.
4. Longitudinal studies evaluating the effectiveness of emerging AI copyright regulations, including the UK's proposed disclosure requirements and the EU's transparency obligations.
5. Forward-looking research on how developments in AI technology (such as few-shot learning, federated training, or synthetic data generation) might affect the legal landscape and require new regulatory approaches.
6. Development of machine-readable licensing standards, establishment of enforceability mechanisms, and evaluation of multiple deployment scenarios.

## Acknowledgments

The author gratefully acknowledges the peers’ contributions:
Ludo Waltman: Supervision, Writing – review & editing
Erna Sattler: Writing – review & editing

## Author’s Contribution

Dmitry Kochetkov: Conceptualization, Investigation, Writing – original draft

## AI Assistance Disclosure

This manuscript was prepared with the assistance of artificial intelligence tools for the following purpose:

- Translation assistance for reading and interpreting source texts (Yandex.Translate (YaGPT-5) and DeepSeek R3)
- Proofreading and linguistic refinement of the final manuscript version (Yandex.Translate (YaGPT-5) and DeepSeek R3)
- Retrieval of legislative acts and legal precedents (Perplexity AI)